# Design and applications of a fluorescent labeling technique for lipid and surfactant preformed vesicles

**Fanny Mousseau\*, Jean-François Berret\* and Evdokia K. Oikonomou\***

*Laboratoire Matière et Systèmes Complexes, UMR 7057 CNRS Université Denis Diderot Paris-VII, Bâtiment Condorcet, 10 rue Alice Domon et Léonie Duquet, 75205 Paris, France*

**Abstract:** Amphiphilic molecules such as surfactants, lipids and block copolymers can be assembled into bilayers and form vesicles. Fluorescent membrane labeling methods require the use of dye molecules that can be inserted in the bilayers at different stages of the synthesis. To our knowledge there is no generalized method for labeling preformed vesicles. Herein we develop a versatile protocol that is suitable to both surfactant and lipid preformed vesicles and requires no separation or purification steps. Based on the lipophilic carbocyanine green dye PKH67, the methodology is assessed on zwitterionic phosphatidylcholine vesicles. To demonstrate its versatility, it is applied to dispersions of anionic or cationic vesicles, such as a drug administrated to premature infants with respiratory distress syndrome, or a vesicle formulation used as a fabric softener for homecare applications. By means of fluorescence microscopy we then visualize the interaction mechanisms of nanoparticles crossing live cell membranes and of surfactant adsorbing on cotton fabric. These results highlight the advantages of a membrane labeling technique that is simple and applicable to a large number of soft matter systems.

## 1 - Introduction

Amphiphilic molecules such as double-tailed surfactants, lipids or block copolymers can assemble in aqueous solutions into bilayers and form various structures including liquid-crystal smectic or cubic phases, multiconnected sponge analogs or closed vesicles. Membrane composition, structure and elastic properties have earned widespread attention because membrane-based colloids are implemented in many industrial and medical applications.[1] In chemistry, vesicles enter into topical home and personal care formulations through their ability to adsorb onto various substrates.[2] In nanomedicine, an important application relates to drug delivery and the use of liposomes as nanocarriers in anti-cancer chemotherapies.[3] Under the impetus of biology, there has been recently a significant increase of novel fluorescence techniques that allow to visualize membranes via optical microscopy, and to follow complex phenomena in real time.[4-7] The interest for fluorescence microscopy has further increased with the recent discovery of super-resolution imaging techniques that allow to go beyond the diffraction limit with optics.[8]

In this context, it is important to control the membrane labeling techniques. For surfactants, the mono- and bilayer labeling takes advantage of fluorescently-tagged amphiphiles[9-11] or solvatochromic surfactant-dyes,[12] the latter having the property to change the emission wavelength depending on their environment. With surfactants however, it is often found that the dye insertion modifies the overall assembly properties, inducing a disruption of micellar or membrane structures. With lipids, the most common labeling methods are based on lipid film hydration[5-6,13-14] or electroformation[5,15] in which a small proportion of fluorescently modified lipids is added beforehand. This technique is suitable for the production of giant uni-lamellar vesicles and has shown excellent results.[5-7,13] The previous methods are however not suitable for living cell membranes or for preformed vesicles. Here, we use the term "preformed vesicles" to describe surfactant or lipid compartments obtained using various chemical or biological processes and which structure must be preserved upon staining. Preformed vesicles are for instance present in cosmetics or homecare formulations.[16-18] In biology they are found in lung fluids or they are the exosomes secreted by live cells associated to regular trafficking functions.[19-21] In the recent literature, we have not found any vesicle labeling protocol that is simple enough and suitable for a broad range of amphiphilic molecules, including surfactants and lipids. After evaluating several biology-based staining methods, we show here an easy and versatile solution to this problem.

In cellular biology, membrane labeling is commonly achieved using lipophilic carbocyanine dyes.[22-23] These dyes intercalate spontaneously and non-covalently into the bilayers thanks to their aliphatic pending chains and rapidly stain the outer cellular membrane through lateral diffusion.[24-25] With time, the dye molecules also spread to intracellular organelles as a result of lipid exchange. Well-known examples are the carbocyanine fluorescent dyes PKH (e.g. PKH26, PKH67) and DiO, DiI, DiD and DiR which cover the visible and infrared emission spectrum.[21-22,26-28] Direct labeling using lipophilic dyes can however lead to significant issues. To speed up the insertion of fluorescent molecules into the membrane, large concentrations of lipophilic molecules and the use of specific solvents are prescribed, leading to further complex separation protocols.[20-21,27] For preformed vesicles, the difficulty is augmented because the vesicles and the non-solubilized dye aggregates have comparable sizes and densities. In such cases, more sophisticated separation techniques based on sucrose gradient or on chromatography are to be implemented.[21]

Herein we develop a versatile labeling membrane protocol that is suitable to both surfactant and lipid preformed vesicles and requires no separation or purification steps. The methodology is based on direct staining using the PKH67 liphophilic green dye (Scheme 1). The protocol design and optimization are first assessed using dipalmitoylphosphatidylcholine (DPPC) preformed vesicles. The DPPC vesicle features such as the size and the charge are thoroughly investigated and found in agreement before and after the labeling treatment. The protocol is then extended to two types of vesicle dispersions: i) an exogenous pulmonary surfactant drug administrated to premature infants with respiratory distress syndrome and ii) a vesicle formulation used as a conditioner in homecare applications. These vesicles differ by their surface charges, the former being anionic and the latter cationic. In the end, we show the benefits of such a staining protocol in two applications, namely the role of pulmonary surfactant in the particle internalization in live cells and the surfactant deposition on cotton fabric.

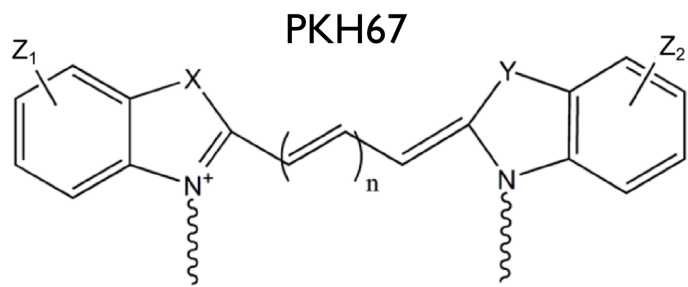

**Scheme 1:** *Chemical structure of PKH67 amphiphilic dye. The molecule is made of the fluorescent head groups $Z_1$ and $Z_2$ emitting in the green (502 nm) and of two aliphatic chains that insert into the bilayer (see Materials and Methods for details).*

## 2 - Results and Discussion

### 2.1 - PKH67 behavior in polar solvents

Dynamic light scattering was used to study the PKH67 dye solubility in solvents such as ethanol, DI-water, PBS phosphate buffer and Diluent C®. In ethanol, no aggregates nor micelles are observed with light scattering indicating a good solubility in this medium. Diluent C® is the commercially available solvent for carbocyanine PKH dyes. It is used primarily to maintain the cell viability whilst maximizing the dye solubility in the culture medium. As Diluent C® components may change the vesicle properties (it contains surfactants, proteins and complexing agents), its use requires post-labeling treatment such as purification and separation steps. Alternative solvents to Diluent C® are water and phosphate buffer saline (PBS), this later being the reference buffer in many biological studies. Figs. 1a-c displays the intensity distributions of 5 μM PKH67 dispersions in DI-water, PBS and Diluent C® respectively. The size distributions are obtained from the deconvolution of the scattered light autocorrelation function and reveal the existence of nanometer (for DI-water and Diluent C®) to micrometer (for PBS) sized aggregates. Obtained by fitting, the continuous lines show maxima at 400 nm, 1.40 μm and 430 nm. The results suggest that the dyes are partially soluble in either of the solvents, a result that is explained by the hydrophobicity of the fluorescent molecule (**Scheme 1**). From the aggregate sizes, it is inferred that the PKH67 solubility in DI-water is comparable to that in Diluent C® and is better than in PBS. Since the membrane labeling efficiency is linked to this solubility, we conclude that DI-water can be an alternative solvent to Diluent C® for staining vesicles.

### 2.2 - Incorporating PKH67 dyes into lipid vesicles

Systematic studies were performed by mixing DPPC vesicles at a concentration of 1 g $L^{-1}$ in DI-water together with dye dispersions of increasing concentrations. We followed the scheme displayed in **Fig. 2** for the sample preparation and increased the PKH67 molar concentration from 0.2 to 5 μM. These concentrations correspond to a dye-to-lipid ratio varying between 1:300 and 1:7000. It was found that at a ratio of 1:1400, the dyes did not alter the vesicular structure and the vesicles were fluorescent. The vesicle properties prior and after labeling were further examined using fluorimetry, light scattering and microscopy.

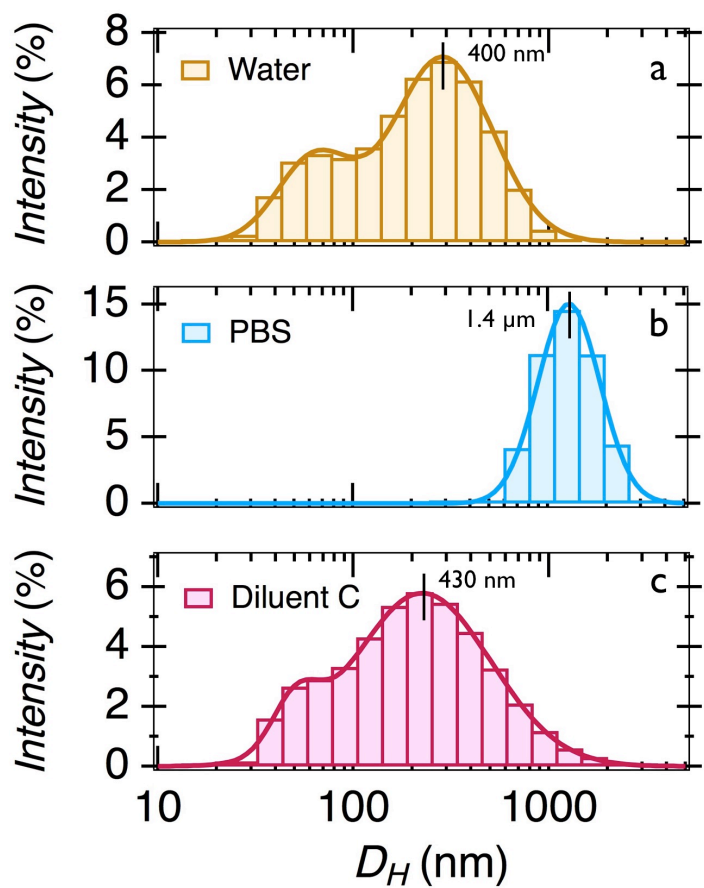

***Figure 1***: *Dynamic light scattering size distributions of 5 μM PKH67 dispersions using **a)** de-ionized water, **b)** PBS and **c)** Diluent C® as a solvent.*

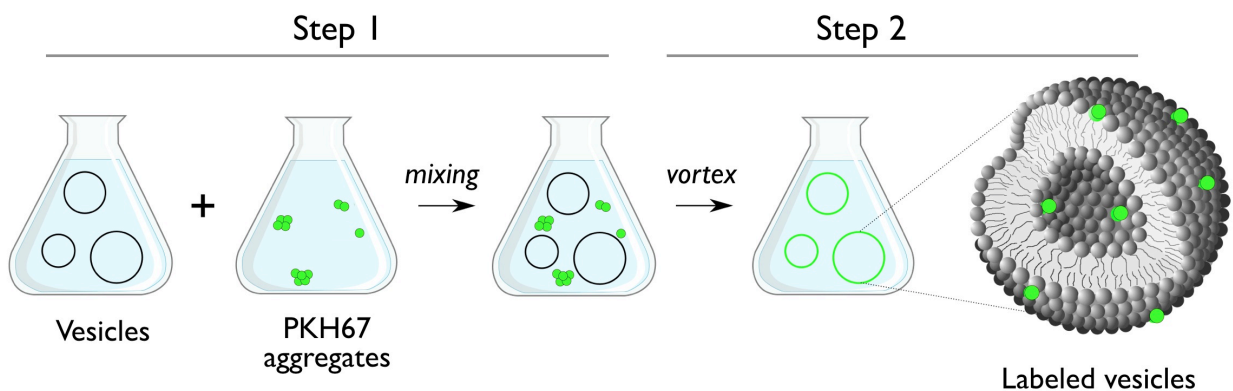

***Figure 2***: *Schematic representation of the protocol used for labeling dipalmitoylphosphatidylcholine (DPPC) vesicles in DI-water.*

**Fig. 3a** shows the fluorescence excitation ($\lambda_{ex} = 490\ nm$) and emission ($\lambda_{em} = 502\ nm$) spectra for a 50 nM PKH67 aqueous suspension and for a PKH67/DPPC dispersion prepared in DI-water as described above. In particular the dye-to-lipid ratio was set at 1:1400. The fluorescence signal observed for DPPC/PKH67 is 40 times higher than that of PKH67 alone at the same concentration. The low fluorescence level found for pure PKH67 dispersion is explained by the phenomenon of aggregation-caused quenching.[29-30] Aggregation-caused quenching takes place when poorly soluble molecules associate into micelles and aggregates, resulting in concentration induced emission quenching. In contrast, the strong signal obtained in the presence of lipids suggests that the dyes have been incorporated in the membranes and are spatially dispersed. There, the distance between dye molecules is estimated at 30 nm, preventing the self-quenching to take place. In both cases the emission occurs at the same wavelength, showing that the presence of DPPC molecules does not alter the PKH67 photophysical properties.

Labeled and unlabeled vesicles were further studied using dynamic light scattering, zeta potential measurements and optical microscopy. Fig. 3b displays the size distribution obtained for a 1 g L$^{-1}$ dispersion with and without PKH67. In the two experiments the distributions peak is around 0.8 - 1 µm. In dilute solutions, the vesicles observed in 60× phase-contrast microscopy are identified as separated and immobile objects adsorbed at the glass surface (Figs. 3c). With fluorescence, the vesicles appear as bright spots co-localized with those observed in phase-contrast (Fig. 3d). An analysis of the merge image (Fig. 3e) leads to the result that more than 80% of the treated vesicles have been labeled and that their size is not modified by the staining. These findings confirm that the above protocol is efficient and does not change the overall vesicle geometry.

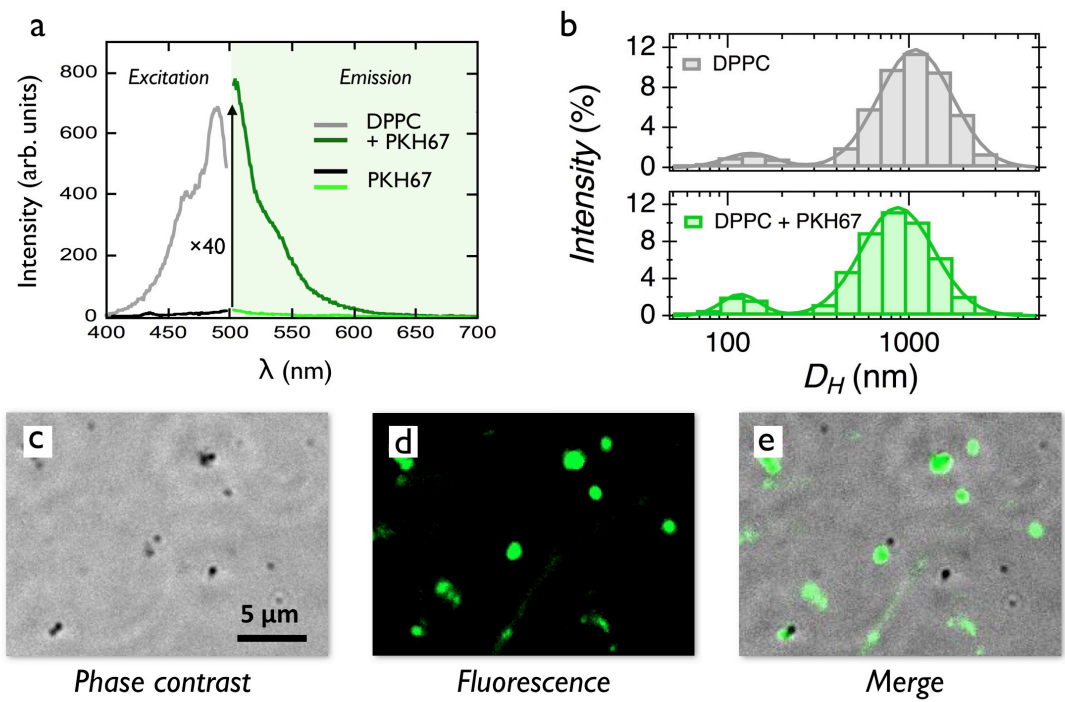

**Figure 3:** a) *Fluorescent excitation and emission spectra of a 50 nM PKH67 aqueous dispersion in presence and in absence of dipalmitoylphosphatidylcholine (DPPC) vesicles. The lipid concentration is 50 mg L$^{-1}$ and the temperature 25 °C. The arrow indicates that with DPPC the fluorescence signal is enhanced by a factor*

*40.* ***b)*** *Dynamic light scattering size distribution of PKH67 labeled and unlabeled DPPC vesicles at 1 g L$^{-1}$.* ***c-d)*** *Phase-contrast and fluorescence optical microscopy images showing DPPC vesicles adsorbed at a PDADMAC coated glass substrate.* ***e)*** *Merge of* ***c)*** *and* ***d)*** *showing co-localization.*

## 2.3 - Labeling biological lipid and industrial surfactant vesicles

The previous method was applied to vesicles of different origin and surface charge. Vesicles made from biological lipids coming from the pulmonary surfactant drug Curosurf® or those made from the commercial double-chain surfactant TEQ were assessed. We recall that Curosurf® is a lung fluid sub­stitute used for the treatment of respiratory distress syndrome.[31-33] Its composition comprises a wide variety of lipids (Scheme 2) which impart an average negative charge to the membrane (Table I). The TEQ surfactant in turn is a quaternary ammonium esterquat entering topical softener formulations. In Supplementary Information S1 and S2 cryo-TEM images confirm that both Curosurf® biological li­pids and TEQ esterquat surfactants associate locally into bilayers with a thickness of 4 - 5 nm and that on a larger scale the bilayers form unilamellar and multivesicular vesicles.[31] Figs. 4a and 4b display the size distributions for PKH67 labeled and unlabeled Curosurf® and TEQ dispersions respectively obtained from light scattering at 1 g L$^{-1}$. In this assay, the Curosurf® vesicle size is controlled *via* extrusion through a 100 nm pore polycarbonate membrane.[34] The data indicate that there is a good superimposition of the distributions for the two dispersions. Electrophoretic mobility experiments show that the zeta potential remains unchanged with and without labeling, the values being around -20 mV and +60 mV for Curosurf® and TEQ respectively (Table I). In Figs. 4c-e and f-h, phase-contrast and fluorescent microscopy images of native non-extruded Curosurf® and TEQ are presented. The figures reveal well-contrasted spherical objects of average size 1 μm. As for DPPC, the vesicles appear as bright spots co-localized with those observed in phase-contrast mode. 85 % of them are fluorescent and their size distribution remain unchanged. These findings confirm that the staining protocol devel­oped for zwitterionic DPPC vesicles can be extended to anionic lipids from biological origin and to cationic surfactants coming from an industrial source. In the next sections, we take advantage of this labeling technique to investigate the interaction mechanisms of nanoparticles crossing live cell mem­branes and of surfactant adsorbing on cotton fabrics.

| Vesicles | $D_H$ (nm) | | $\zeta$ (mV) | |
|---|---|---|---|---|
| | Native | Fluorescent | Native | Fluorescent |
| DPPC | 1350 | 1080 | 0 | 0 |
| Pulmonary surfactant substitute Curosurf®* | 130 | 135 | - 20 | - 18 |
| Double-tail esterquat surfactant TEQ | 550 | 530 | + 60 | + 58 |

***Table I***: *Hydrodynamic diameter ($D_H$) and zeta potential ($\zeta$) of the native and fluorescent DPPC, Curosurf® and TEQ vesicles. The star (\*) indicates that the vesicles are extruded with a 100 nm pore polycarbonate membrane.*

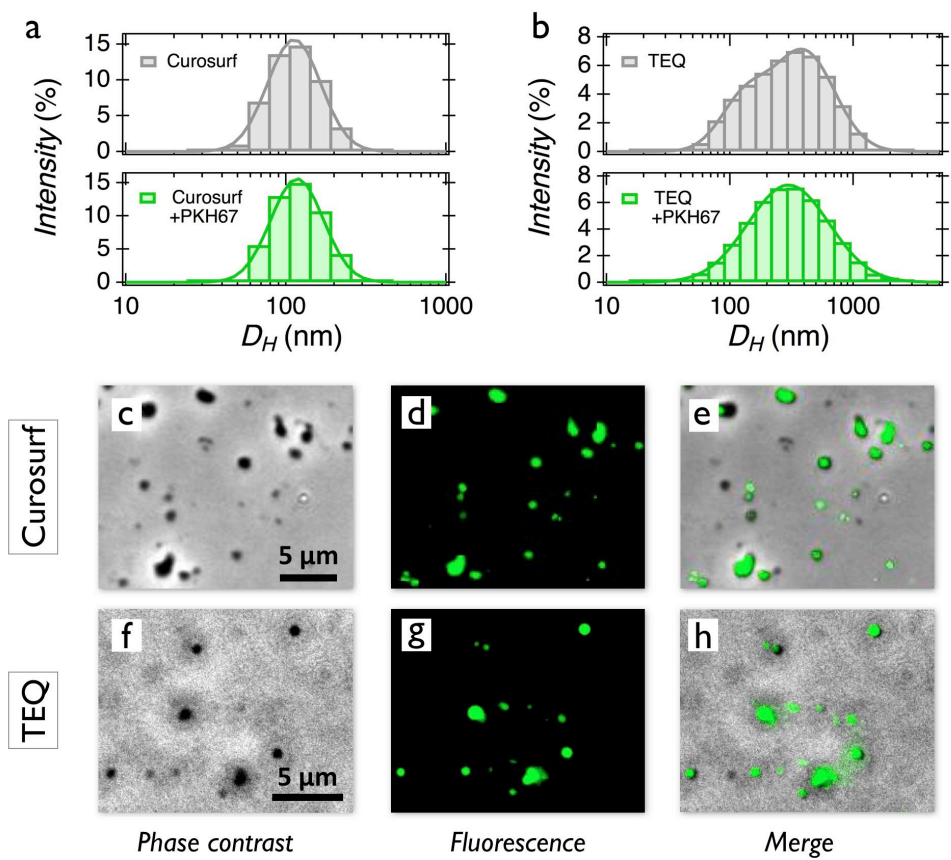

**Figure 4: a)** *Dynamic light scattering size distribution of 1 g L⁻¹ extruded Curosurf® vesicles with and without PKH67 dyes. The extrusion was made using a 100 nm pore polycarbonate membrane.* **b)** *Similar to* **a)** *for native TEQ vesicles.* **c-d)** *Phase-contrast and fluorescence optical microscopy showing non-extruded Curosurf® vesicles.* **e)** *Merge of the patterns found in phase-contrast and fluorescence.* **f-h)** *Same as* **c-e)** *for TEQ quaternary ammonium esterquat surfactant.*

## 2.4 - Role of the pulmonary surfactant in the cellular uptake of nanoparticles

When inhaled, nanoparticles are able to reach the respiratory zone and enter in contact with the alveolar region.[35-37] Various scenarios of nanoparticles passing from the alveolar spaces towards the blood circulation have been examined and in some instances translocation has been demonstrated.[37-38] Our research group has recently explored the behavior of 50 nm particles and of surfactant substitutes in controlled physico-chemical conditions. Three types of surfactant mimetics, including the exogenous substitute Curosurf® were assessed together with aluminum oxide, silicon dioxide and latex nanoparticles.[34] The result that emerged from this survey (particles were selected to display different shapes and surface charge densities[39]) was the observation of the spontaneous nanoparticle-vesicle aggregation induced by Coulomb attraction. The aggregation was strongly enhanced for oppositely charged species. Labeling pulmonary surfactant vesicles, as shown previously represents hence an opportunity to further study the aggregate structure. Fig. 5a displays a high magnification view of a silica/ Curosurf® aggregate observed under phase-contrast (Fig. 5a1), green (Fig. 5a2) and red (Fig. 5a3) emission. In this assay, the particles are aminated and fluoresce in the orange-red at 590 nm due to rhodamine molecules present in the silica structure. The experimental conditions are a total concentration of 1 g

L⁻¹, a volumetric lipid-to-nanoparticle ratio of 2 and a temperature of 37 °C.[34] The merge image of Figs. 5a2 and 5a3 exhibits an excellent superimposition of the green and red channels over the entire object, indicating that the aggregates contain both fluorescent species (Fig. 5a4). These results, together with those of Supplementary Information S3 ascertain that the aggregates are made of vesicles and particles and that both are intertwined at the micron scale (Fig. 5b).

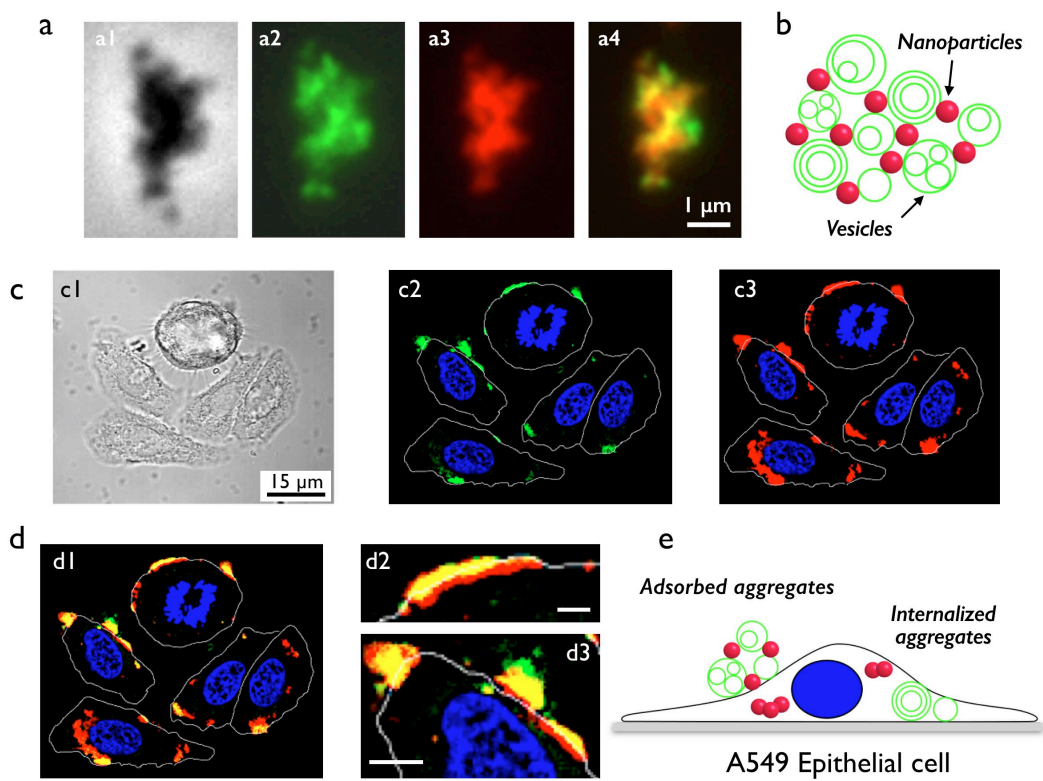

*Figure 5: **a)** A silica nanoparticle/ Curosurf® aggregate observed under phase-contrast (**a1**), green (**a2**) and red (**a3**) illumination. The silica fluoresce in the orange-red at 590 nm and the vesicles are labeled with a green fluorescent dye (PKH67) emitting at 502 nm. The merge image of the red and green signals is shown in **a4**. **b)** Schematic representation of a nanoparticle/ Curosurf® aggregate derived from optical microscopy. **c)** Bright field (**c1**) and confocal microscopy (**c2,c3**) of A549 alveolar epithelial cells incubated with silica nanoparticle/Curosurf® aggregates at concentrations 0.10/0.015 g L⁻¹. The nuclei are labeled in blue with DAPI. The white lines are the cell contours. **d)** Three-color merge images showing cells (**d1**), plasma membrane (**d2**) and cytoplasm (**d3**). The bars in **d2** and **d3** are 2 μm and 4 μm respectively. **e)** Schematic representation of a cell with adsorbed and internalized silica nanoparticles.*

In the alveolar spaces, inhaled nanoparticles enter first in contact with the lipid monolayer at the air-liquid interface, and then with the hypophase.[40-41] This interaction leads to an aggregate formation comparable to the one discussed previously and is susceptible to modify the fate of the nanoparticles towards the alveolar cells located beneath the hypophase. In particular, the lipids are expected to mitigate the protein adsorption.[42] Here, the double labeling permits to visualize the vesicle and particle localization following the cellular uptake. A549 lung epithelial carcinoma cells were incubated for 4 h with the nanoparticle-vesicle aggregates shown in Fig. 5a. Due to their size and density, the aggregates sedimented at the bottom of the Petri dish[36,43] and enter in contact with the cell layer. Fig. 5c exhibits confocal images of a cluster of five cells in bright field (Fig. 5c1) and in green and red fluorescence (Figs. 5c2 and 5c3 respectively). The nuclei are stained in blue with DAPI. Fig. 5d exhibits the merge image of the three fluorescent signals, together with close-up views of the plasma membrane and of the cytoplasm. These images demonstrate that in the presence of lipid membranes, the particles are internalized in the cells in a significant amount (see Supplementary Information S4 for additional data). In this respect, the vesicular membrane seems not to have a protective role towards cells, as it is the case with supported lipid bilayers.[31] Yellow patches associated with nanoparticle-vesicle co-localization are essentially visible at the outer cellular range, but not internally or in a lesser amount. This suggests that after passing through the outer lipid barrier a separation of the two components occurs and that the preformed aggregates break. Fig. 5e illustrates schematically this process. To our knowledge, it is the first time that assays using membranes of biological origin reveal a separation mechanism for the organic and inorganic species in cells.

**2.5 - Double-chain surfactant deposition on cotton fibers**

Recently, we show that PKH67 labeled vesicles made from TEQ double-tailed surfactant could be also used to decipher the interaction mechanism with cellulose nanocrystals.[17] Obtained from natural cellulose,[44] cellulose nanocrystals are rod-shaped particles in the form of 200 nm laths. Being negatively charged, these anisotropic particles interact strongly with the TEQ cationic vesicles and form mixed aggregates comparable to those obtained with the silica nanoparticles and discussed in the previous section. Examples are given in Supplementary Information S5.

To evaluate the performances of a home care product such as a fabric softener, it is more convenient to use actual fabric materials for deposition assessment. Pertaining to the deposition on cotton, important issues are the amount of adsorbed surfactants and the morphology of the deposited layer, questions to which labeled vesicles could provide an answer.[45-48] In a typical experiment, a 1×1 cm$^2$ piece of woven cotton fabric is treated with a dispersion containing TEQ fluorescent vesicles at 1 g L$^{-1}$ during 10 minutes and then rinse with DI-water. The PKH67 concentration was set at 6 μM for these assays, so slightly higher than for DPPC and Curosurf® vesicles. Fig. 6a displays a phase-contrast microcopy view of a treated fabric submerged in water. There, the 200 μm sized yarns are recognizable and separated by large voids. Additional SEM data confirm the woven structure (Supplementary Information S6). In the emission mode (Fig. 6b), the fibers exhibit an intense fluorescence signal, illustrating that TEQ surfactants have been adsorbed on the cellulose substrate in a significant manner. The bright spots (arrows) are assigned to large TEQ vesicles coating the cellulose substrate. The vesicle adsorption is explained by Coulomb attraction between the cationic vesicles and the anionic fibers. Control experiments performed with unlabeled surfactant show that in the present conditions the fibers are not fluorescent and appear dark (Supplementary Information S7). An analysis of the distribution of the fluorescence signal along the fiber cross section shows an enhanced signal on the edges, the intensity showing an M-shape signal (Fig. 6c, 6d and inset in Fig. 6c). Such M-shape signals are characteristic from fluorescence emission arising from the fluorescent surfaces or interfaces.[6,49] Concerning the nature of the absorbed layer, it could come either from supported lipid bilayers or from a supported vesicle layer. In the first scenario, the fibers would be coated with a single or with a stack of surfactant bilayers, whereas in the second the vesicles remain intact and stick to the cellulose fibers. Figs. 6b and 6d actually suggest that both scenarios are here plausible. These two types of surfactant deposition are schematically presented in Fig. 6e. In conclusion, the present technique allows a direct visualization of surfactant adsorbed on cotton and should permit to assess the formulation deposition performance.

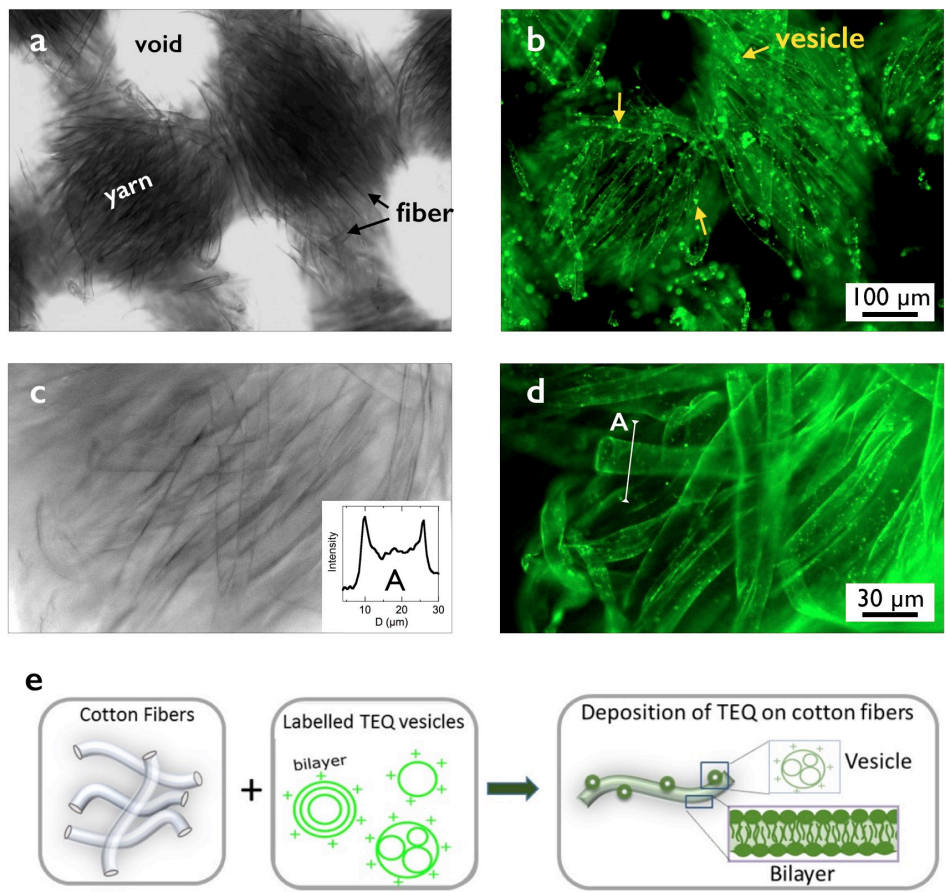

***Figure 6****: Phase-contrast (**a,c**) and fluorescence (**b,d**) microscopy images of a woven cotton fabric treated with fluorescent vesicles. Magnification are 20× for **a** and **b**, and 40× for **c** and **d**. In **b)** fluorescent vesicles are indicated with yellow arrows. Inset in **c)**: fluorescence profile along the line A depicted in **d)**. The fluorescence maxima are ascribed to the fiber edges. **e)** Schematics of surfactant deposition on cotton fibers assuming either supported lipid bilayer or supported vesicle layer models.*

## 3 - Conclusion

In this work we report a fluorescent labeling technique based on the spontaneous insertion of lipophilic molecules into membranes made either from surfactants or lipids. The protocol is derived from methods developed for live cell imaging. Thanks to an accurate dosage of the dye molecules in solution, here the PKH67 lipophilic green fluorophore, issues related the removal of the excess dyes or to the use of a specific solvent are avoided. The PKH67 concentration is namely adjusted to that of the lipids in a way that the fluorescence signal remains high and that at the micron scale the vesicular structures are not altered by the dye insertion. Optimum ratios are in the range 1 dye molecule per 1400 lipids. The vesicular resilience upon dye addition is evaluated on DPPC vesicles using light scattering, zeta potential and fluorescent microscopy. With the above dye-to-lipid ratio, we insure moreover that phenomena such as aggregation-caused quenching do not occur and mitigate the fluorescence intensity. As anticipated, we also found that separation and centrifugation techniques usually necessary to remove the dye excess are here not required. To demonstrate its versatility, the protocol is applied to two other vesicular systems, a lipid drug administrated to premature infants with respiratory distress syndrome (Curosurf®) and a surfactant formulation used as fabric conditioner. In Curosurf® the vesicles are negatively charged, whereas in the surfactant formulation they are positively charged. Interestingly, the amounts of dye added to optimize the fluorescence signals depend little on the system or on the electrostatic charge, around 1 μM for a lipid or surfactant solution at 1 g L$^{-1}$. To complete this study, we illustrate the benefits of this labeling method in evaluating the internalization of silica nanoparticles in cells after their interactions with Curosurf® vesicles. With the fabric softener formulation, we also investigate the deposition of surfactant on cotton fibers. In both examples, we are able to visualize using fluorescence microscopy the fate of the vesicles after their interactions with live cells or with cotton. These results allow us to establish the interaction patterns specific to each system and to assign the relevant nanoscale interaction mechanisms thanks to fluorescence microscopy.

# 4 - Materials and Methods

## 4.1 - PKH67 fluorescent dye

Developed for live cells, PKH linkers are lipophilic molecules that incorporate spontaneously into the lipid bilayer. Their photophysical properties cover the visible and infrared emission spectrum and the molecules are described as non-toxic effects towards cells.[20,22,26] The 1 mM PKH67 solution in ethanol was purchased from Sigma-Aldrich and used as received. The molecule is made of two fluorescent head groups and two aliphatic chains. Its chemical structure as provided by the supplier is shown in Scheme 1. PKH67 fluoresces in the green, its excitation and emission wavelengths being at 410 nm and 502 nm respectively. More information can be found in the website in Supplementary Information S5 and S7.

## 4.2 - Phospholipids and surfactants

Dipalmitoylphosphatidylcholine (DPPC, Scheme 2a) was purchased from Sigma-Aldrich. It is a phosphatidylcholine lipid with a gel-to-fluid phase transition temperature of 41 °C.[50] For vesicle preparation, DPPC was initially dissolved in methanol at 10 g L$^{-1}$. After a 30 min solvent evaporation at low pressure and 60 °C, a lipid film was formed on the glass surface. It was hydrated with the addition of deionized (DI) water at 60 °C and stirred at atmospheric pressure for another 30 minutes. DI-water was added to obtain a 1 g L$^{-1}$ dispersion. In these conditions, DPPC was found to form uni- and multilamellar vesicles.

Curosurf®, also called Poractant Alfa (*Chiesi Pharmaceuticals*, Parma, Italy) is an extract of whole mince of porcine lung tissue purified by column chromatography.[33,51] It is indicated for the rescue treatment of respiratory distress syndrome in premature infants and is administered intratracheally at a dose of 200 mg per kilogram. Curosurf® was kindly provided by Dr. Mostafa Mokhtari and his team from the neonatal service at Hospital Kremlin-Bicêtre, Val-de-Marne, France. Curosurf® is produced as a 80 g L$^{-1}$ phospholipid and protein suspension where the phospholipids are assembled in the form of multi-lamellar anionic vesicles. It contains, among others, phosphatidylcholine lipids, sphingomyelin, phosphatidylglycerol and the membrane proteins SP-B and SP-C.[32,34,40,52] Some of the lipids present in Curosurf® are displayed in Scheme 2b (Supplementary Information Table S1). According to the manufacturer, the pH of Curosurf® is adjusted with sodium bicarbonate at pH 6.2.[53] In the conditions used in this work, Curosurf® is organized in anionic vesicles with a large size distribution (50 nm − 2 μm), as shown by cryogenic transmission electron microscopy (cryo-TEM[54]). The gel-to-fluid transition of Curosurf® membranes was measured and found at $T_M$ = 29.5 °C (Supplementary Information S8).

The esterquat cationic surfactant ethanaminium, 2-hydroxyN,N-bis(2-hydroxyethyl)-N-methyl-esters with saturated and unsaturated C16-18 aliphatic chains (Scheme 2c), abbreviated TEQ in the following was provided by Solvay®. Its gel-to-fluid phase transition temperature is observed at 60 °C and the counterions associated with the quaternized amines are methyl sulfate anions.[17] TEQ is a cationic surfactant widely used in the industry as the main component of hair and fabric conditioners. It forms unilamellar and multi-vesicular vesicles in water, as shown by cryo-TEM in Supplementary Information S2.[17]

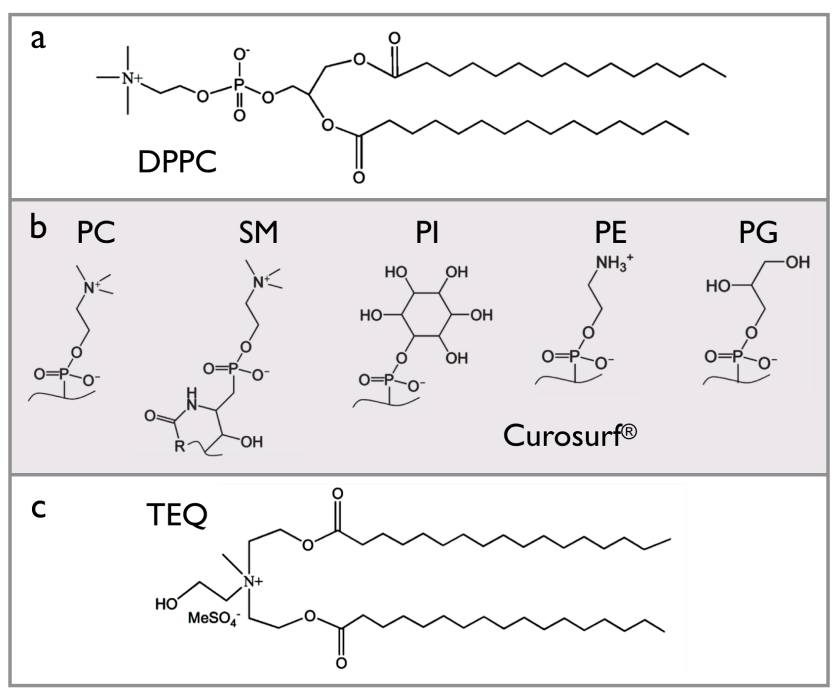

**Scheme 2:** *Chemical formulae of a) dipalmitoylphosphatidylcholine (DPPC), b) most common lipids present in Curosurf® formulation and c) ethanaminium, 2-hydroxyN,N-bis(2-hydroxyethyl)-N-methyl-esters abbreviated TEQ. Curosurf® contains zwitterionic (phosphatidylcholine (PC), sphingomyelin (SM), phosphatidylethanolamine (PE)) and anionic lipids (phosphatidylinositol (PI), phosphatidylglycerol (PG)). Its composition is available in Supplementary Information Table S1.*

### 4.3 - Silica nanoparticles

The positively charged silica particles were synthetized using the Stöber synthesis.[31,55] Briefly, fluorescent silica seeds were prepared in three steps. Rhodamine derivative rhodamine red c2 maleimide (Aldrich) was first covalently bound to silica precursor (3-mercaptopropyl)-trimethoxysilane (MPS, Aldrich). The rhodamine–MPS compound was then mixed with a tetraethyl orthosilicate silica precursor (TEOS, Aldrich) and the Stöber synthesis was performed. With this approach the dyes were covalently bound to the silica matrix. In a third step, a non-fluorescent silica shell was grown with TEOS to increase the particle size and prevent leakage of the dyes out of the particles. Functionalization by amine groups was then performed, resulting in a positively charged coating.[55] Aminated silica was synthesized at 40 g L$^{-1}$ and diluted with DI-water at pH 5. The hydrodynamic and geometric diameters were determined as $D_H = 60$ nm and $D_0 = 41.2$ nm. The fluorescence properties were characterized using a Cary Eclipse fluorimeter (Agilent), with leading excitation and emission peaks at 572 nm and 590 nm, respectively. A particle identity card displaying UV-Vis spectrometry, fluorimetry, TEM and light scattering data is provided in Supplementary Information S9.

### 4.4 - Cellulose fibers

1×1 cm$^2$ woven cotton fabrics were used to investigate the softener surfactant deposition onto fibers. The fabrics are made of cotton yarns of diameters $300 \pm 100$ μm, each of them being constituted of approximately 20 cotton fibers of diameter 10 to 20 μm. A representative SEM image of the woven cotton fabric used in this work is provided in Supplementary information S6. All fabrics were first treated with DI-water for 10 min and dried at 35 °C under air circulation before use.

### 4.5 - Vesicle labeling

The 1 mM commercial PKH67 stock solution in ethanol was first diluted 10 times using pure ethanol. 2 μl of the diluted stock was then mixed rapidly with 100 μl of DI-water and vortexed for 10 s. It was then added to a 100 μl of a DPPC or Curosurf® vesicle aqueous dispersion at the concentration of 2 g L$^{-1}$. The final dye concentration was then 1 μM. Rapid vortexing for *ca.* 10 s was followed to ensure the dye insertion in the lipid membranes. The dispersion was let to rest in the dark for 10 minutes. For labelling TEQ vesicles, the dye concentration used was 6 times larger than the previous one. In particular, 2 μl of the PKH67 stock solution were rapidly mixed with 100 μl of DI-water and vortexed for *ca.* 10 s. This solution was then added to 100 μl of a TEQ vesicle dispersion at the concentration of 2 g L$^{-1}$ and the mixture was vortexed.

### 4.6 - Sample preparation

To evaluate the PKH67 solubility, solutions were prepared at 5 μM in different solvents, such as ethanol, PBS, Diluent C® and water. Diluent C® is a commercial product developed for cell labeling. It contains glycerol (50%), Triton X100 (0.15 wt. %), bovine serum albumin (200 mg L$^{-1}$), dithiothreitol (1 mM), EDTA (0.1 mM), NaCl (250 mM) and Tris-HCL (10 mM). Pertaining to the vesicles, Curosurf®, DPPC and TEQ dispersions were prepared at room temperature using DI-water and a concentration of 2 g L$^{-1}$. Before use, the dispersions were characterized by dynamic light scattering and ζ-potential measurements. In some cases, the dispersions were extruded through a polycarbonate membrane following a protocol described in Ref.[34] to reduce the vesicle size dispersity. To study the interaction of silica nanoparticles with lipids, we use the method of continuous variation developed by Paul Job to determine the stoichiometry of binding (macro)molecular species in solutions.[56] The method is here combined with static and dynamic light scattering, leading to Job scattering plots.[17,34,53,57] In brief, stock solutions (1 mL) were prepared in the same conditions of pH and concentration (1 g L$^{-1}$) and mixed at the volumetric ratios $X = V_{LP}/V_{NP} = 2$ where $V_{LP}$ and $V_{NP}$ are the volumes of the surfactant and particle solutions. The particle and lipid concentrations (0.66 and 0.33 g L$^{-1}$, respectively) correspond to identical surface area concentrations in membrane and in silica.[34]

### 4.7 - Deposition of labeled TEQ vesicles on cotton fabrics

10 µl of PKH67 were dispersed in 1500 µl of DI-water and mixed rapidly with vortex. This mixture was then added in 1500 µl of TEQ prepared at 1 g L⁻¹ and mixed rapidly with the vortex. 0.03 g of cotton fabrics cut in small pieces were immersed for 10 minutes in this dispersion.

## 4.8 - Cell culture

Adenocarcinomic human alveolar epithelial cells A549 (ATCC reference CCL-185™) were grown in T75-flasks in DMEM supplemented with 10% fetal bovine serum (FBS) and 1% penicillin/strepto-mycin. Exponentially growing cultures were maintained in a humidified atmosphere of 5% $CO_2$ and 95% air at 37 °C. When the cell confluence reached 80%, cell cultures were passaged using trypsin–EDTA. All products used in cell culture came from Gibco, Life Technologies.

## 4.9 - Light scattering

The scattering intensity $I_S$ and hydrodynamic diameter $D_H$ were measured using the Zetasizer NanoZS spectrometer (Malvern Instruments, Worcestershore, UK). A 4 mW He−Ne laser beam ($\lambda$ = 633 nm) is used to illuminate the sample dispersion, and the scattered intensity is collected at a scattering angle of 173°. The second-order autocorrelation function $g^{(2)}(t)$ is analyzed using the cumulant and CONTIN algorithms to determine the average diffusion coefficient $D_C$ of the scatterers. The hydrodynamic diameter is then calculated according to the Stokes-Einstein relation, $D_H = k_B T / 3\pi\eta D_C$, where $k_B$ is the Boltzmann constant, $T$ the temperature and $\eta$ the solvent viscosity. Measurements were performed in triplicate at 25 °C after an equilibration time of 120 s.

## 4.10 - Zeta potential

Laser Doppler velocimetry (Zetasizer, Malvern Instruments, Worcestershore, UK) using the phase analysis light scattering mode and a detection at an angle of 16 degrees was performed to determine the electrophoretic mobility and zeta potential of the different studied dispersions. Measurements were performed in triplicate at 25 °C, after 120 s of thermal equilibration.

## 4.11 - Phase-contrast and fluorescence optical microscopy

Images were acquired on an IX73 inverted microscope (Olympus) equipped with a 60× objective. An Exi Blue camera (QImaging) and Metaview software (Universal Imaging Inc.) were used as the acquisition system. The illumination system "Illuminateur XCite Microscope" produced a white light, filtered for observing a green signal in fluorescence (excitation filter at 470 nm - bandwidth 40 nm and emission filter at 525 nm - bandwidth 50 nm). Thirty microliters of the vesicles dispersion (DPPC, Curosurf® or TEQ) or nanoparticles / vesicles dispersions (Silica particles / Curosurf® or nanocellulose / TEQ) were deposited on a glass plate and sealed into a Gene Frame dual adhesive system (Abgene/ Advanced Biotech). In case of DPPC and Curosurf® vesicles, the glass slides were coated using a cationic polymer (PDADMAC) to improve their adhesion properties based on electrostatic interaction. Images were digitized and treated by the ImageJ software and plugins (http://rsbweb.nih.gov/ij/).

## 4.12 - Confocal microscopy

Images were acquired on an LSM 710 microscope (Zeiss) equipped with a 40× immersion objective and a temperature and carbon dioxide content controller. The A549 cells (ATTC) were first seeded on a glass slide in a 6 wells plate at 200 000 cells per well. After 48 hours of growth in complete white DMEM, the cells were rinsed with PBS and incubated 4 h at 37 °C with a silica- Curosurf® dispersion.

The concentrations are 0.1 g $L^{-1}$ for the particles and 0.015 g $L^{-1}$ for Curosurf® in white DMEM. Cells were then fixed with PFA and nuclei were stained in blue with DAPI. Finally, slides were sealed into a Gene Frame dual adhesive system (Abgene/Advanced Biotech). Images were digitized and treated by the ImageJ software and plugins (http://rsbweb.nih.gov/ij/).

### 4.13 - Absorbance and fluorescence measurements

A UV-visible spectrometer (SmartSpecPlus from BioRad) was used to measure the absorbance of PKH67 aqueous dispersions at 0.05 and 2 µM. The fluorescence properties of PKH67 at 0.05 and 2 µM and of DPPC/PKH67 at 0.05 g $L^{-1}$/0.05 µM were characterized using a Cary Eclipse fluorimeter (Agilent) with PMT at 860, 980 and 860 respectively.

### Conflict of interest

The authors declare no conflict of interest

# Supplementary information

The Supporting Information is available free of charge on the ACS Publications website at DOI: 10.1021/acsomega.XXX

Cryo-TEM images of Curosurf® vesicles, CryoTEM images of TEQ vesicles and analysis, Fluorescence optical microscopy images of Silica – Curosurf® aggregates, Confocal image of silica nanoparticles in cells, Phase contrast and fluorescence microscopy of mixed vesicle-cellulose nanocrystal aggregates, Scanning Electron Microscopy (SEM) of cotton fabric, Control experiments for surfactant deposition on cotton fabric, Differential scanning calorimetry of Curosurf®, Silica particle identity card, Curosurf® composition in lipids and proteins.

# Acknowledgement

We thank Armelle Baeza-Squiban, Victor Baldim, Mélody Merle, Mostafa Mokhtari, Jesus Perez-Gil, Chloé Puisney, Milad Radiom for fruitful discussions. We also acknowledge the support from Galder Cristobal and Jean-Christophe Castaing from Solvay during the research program on novel softener formulations. ANR (Agence Nationale de la Recherche) and CGI (Commissariat à l'Investissement d'Avenir) are gratefully acknowledged for their financial support of this work through Labex SEAM (Science and Engineering for Advanced Materials and devices) ANR 11 LABX 086, ANR 11 IDEX 05 02. We also thank Stéphane Mornet from the Institut de Chimie de la Matière Condensée de Bordeaux (Université Bordeaux 1) for the synthesis of the aminated fluorescent silica nanoparticles. We acknowledge the ImagoSeine facility (Xavier Baudin, Jacques Monod Institute, Paris, France), and the France BioImaging infrastructure supported by the French National Research Agency (ANR-10-INSB-04, « Investments for the future »). This research was supported in part by the Agence Nationale de la Recherche under the contract ANR-13-BS08-0015 (PANORAMA), ANR-12-CHEX-0011 (PULMONANO), ANR-15-CE18-0024-01 (ICONS), ANR-17-CE09-0017 (AlveolusMimics) and by Solvay.

# TOC Image

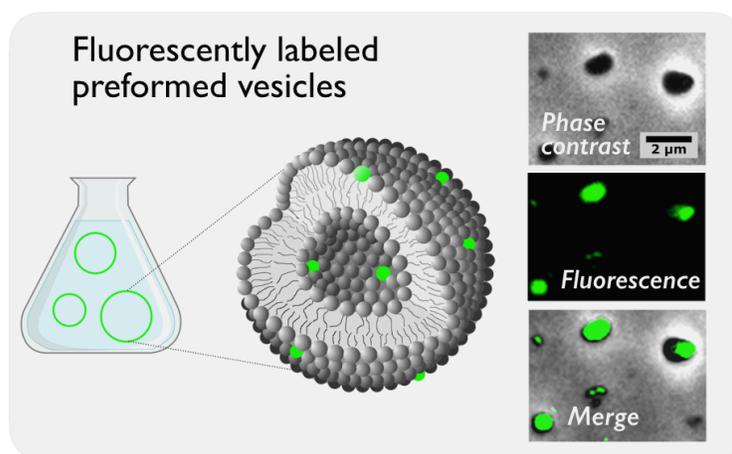